# Integration of IoT- AI powered local weather forecasting: A Game-Changer for Agriculture

**Suman Kumar Das[1], Pujyasmita Nayak[2]**

[1]Department of Mechanical and Industrial Engineering, University of Brescia, Italy

Email: s.das001@unibs.it

[2]Meteorological Centre, India Meteorological Department, Bhubaneswar, Odisha

**Abstract**

The dynamic environment context necessitates harnessing digital technologies, including artificial intelligence and the Internet of Things, to supply high-resolution, real-time meteorological data to support agricultural decision-making and improve overall farm productivity and sustainability. This study investigates the potential application of various AI-powered, IoT-based, low-cost platforms for local weather forecasting to enable smart farming. Despite the increasing demand for this topic, a few promising studies have explored this area. This paper developed a conceptual research framework based on a systematic review of relevant literature and employed a case study method to validate the framework. The framework comprised five key components: the Data Acquisition Layer, Data Storage Layer, Data Processing Layer, Application Layer, and Decision-Making Layer. This paper contributes to the literature by exploring the integration of AI-ML and IoT techniques for weather prediction tasks to support agriculture, and the incorporation of IoT technologies that provide real-time, high-resolution meteorological data, representing a step forward. Furthermore, this paper discusses key research gaps, such as the significant obstacles impeding the adoption of AI in agriculture and local weather forecasting, including the lack of straightforward solutions and the lack of digital skills among farmers, particularly those in rural areas. Further empirical research is needed to enhance the existing frameworks and address these challenges.

## 1. Introduction

The escalating complexity of contemporary agriculture, coupled with the intensifying effects of climate change, underscores the critical need for localized weather forecasting. By leveraging scientific principles and technological innovations, precise and timely weather predictions can equip farmers with the ability to make informed decisions, optimize resource utilization, and bolster agricultural productivity amidst the challenges posed by unpredictable weather patterns. With the help of comprehensive and situation-specific local forecast systems, farmers and other stakeholders can plan and make decisions for the upcoming farming season. An automated weather station with sensors eliminates the need for manual data collection on atmospheric variables like humidity, pressure, temperature, precipitation, and wind. These stations provide weather summaries or forecasts based on the sensor readings. For farmers, measuring the weather parameters in the agriculture zone is crucial because it allows them to plan their farms and supply chain based on the weather. Weather forecasting techniques have been transformed by the integration of artificial intelligence (AI) and the internet of things (IoT), offering greater accuracy and a better understanding of weather patterns. The Internet of Things, a sensor technology, collects and monitors small changes in various parameters through continuous monitoring. This allows for the collection of more accurate and comprehensive data. This wealth of data is essential for running complex AI-based models and algorithms, which require high-quality and substantial quantities of data to effectively process information and accurately predict output variables and outcomes [1]. Hence, to increase the precision, and effectiveness of weather monitoring, IoT-based weather stations are utilizing advanced technologies like artificial intelligence, machine learning, and block chain. Agricultural practices are undergoing a paradigm shift due to artificial intelligence (AI) that include weather forecasting, crop monitoring, soil management, pest identification, weed control, disease management, and more [2]. Integrating climate data with current technology and optimal farming practices is essential. Small-scale farmers require location-specific and crop-specific information that can be readily implemented to improve their agricultural practices [3]. The Internet of Things facilitates advanced device communication and data sharing, primarily through Wi-Fi-based wireless sensor networks. Machine learning techniques can analyze training data to classify and predict information, even from unfamiliar datasets. Emerging low-power wide-area network technologies, such as Sigfox, LoRaWAN, and NB-IoT, are advancing IoT applications in agriculture. Integrating cutting-edge technologies, including machine learning and artificial intelligence, could enhance the accuracy and efficiency of weather monitoring. Additionally, the environmental implications of IoT-based weather stations warrant further investigation.

However, very few academic papers have described a comprehensive framework for integrating technology and technological architecture to fully understand how the integration of these elements takes place, particularly in the context of local weather forecasting and its application to farming operations. Due to this gap, this paper proposes a framework that integrating IoT, AI, and machine learning to improve local weather forecasting for climate-resilient agriculture. This paper provides a comprehensive perspective of AI-powered, IoT-based technologies for localized weather forecasting to support smart farming and enable farmers to adapt to climate change impacts. The proposed framework leverages real-time, heterogeneous meteorological data to empower farmers with accurate weather predictions, optimize resource allocation, and enhance agricultural productivity.

This article is organized as follows. Section 2 provides the research methodology and provides Current IoT-AI research for Enhancing local weather forecast, Section 3 presented framework and Potential Technical solution, Section 4 illustrated the case study. Finally, the article concludes with a summary of findings and insights along with limitations and future direction.

## 2. Related work

### 2.1 Research Methodology

To develop local weather forecasts, the study integrates IoT and AI with agro meteorology. The research process followed narrative literature review and case study method. The researchers searched the Scopus database using relevant keywords related to artificial intelligence, machine learning, and deep learning in the context of agriculture and local weather forecasting. Based on these keywords and their area of interest, the researchers selected appropriate articles for their study. The inclusion criteria focused on papers demonstrating a structured relationship between digital technologies and relevant local weather forecasting. Articles that were too narrow in scope, concentrating on specific strategies like only single technology or only on local weather forecast, were excluded if they did not offer a broad range of technological context or lacked clarity in presenting frameworks or organizing principles. The research process involved conducting a thorough literature review to identify key themes and categorizations relevant to the research topic. This comprehensive literature review aimed to pinpoint essential themes and categorizations, gathering insights from academic sources. The gathered information was then synthesized to construct an integrated framework that likely incorporated various concepts, theories, and findings to provide an organized framework guiding research and analysis. Finally, the framework was validated and refined using practical examples from case studies or real-world applications, ensuring its applicability and relevance in practical scenarios.

## 2.2 Current IoT-AI research for Enhancing local weather forecast.

In current scenario, researchers have developed an innovative system that combines internet-of-things (IoT) and artificial intelligence (AI) technologies to provide localized weather forecasts, supporting climate-conscious agricultural practices as shown in Table. 1. The widespread deployment of low-cost, energy efficient IoT sensor nodes within farmlands has enabled the collection of detailed data on crop conditions and climate factors [4].

Different IoT-driven smart farming solutions have been extensively researched and developed. For example, Faid et al. proposed a low-cost IoT system that uses AI algorithms to enable energy-efficient smart farming. The system monitors pressure, temperature, and humidity to provide real-time forecasts. Heterogeneous nodes and base stations are essential to the proposed system, with sensors collecting data and sending it to the base stations for wireless analysis and backup to the cloud. This research used a Raspberry Pi microcomputer, capable of multitasking, and employed recurrent neural networks and long- and short-term memory models for data analysis [5]. Other researchers have also developed IoT-based weather prediction systems. Parashar proposed an Arduino-based system with eight sensors, analyzing the data using MATLAB and multiple linear regressions on Jupyter Lab [6]. Sarkar et al. [7] introduced an approach combining machine learning and IoT for precise weather forecast analysis and affordable solutions, using IoT sensors to obtain real-time data on humidity, temperature, pressure, and precipitation. Singh developed a weather station using ThingSpeak, a free IoT cloud, to store and visualize the meteorological data obtained from sensors. Math and Dharwadkar proposed an automated weather station integrated with the Internet of Things for precision agriculture in India. This system optimizes resource usage, notifies farmers about real-time crop data, and monitors the data itself. Their low-cost setup utilizes Thing-Speak for data analysis and visualization, Arduino for weather station configuration, and a microcontroller for data transmission. The system encompasses components for data collection, storage, and analysis. However, traditional rainfall prediction and weather forecasting models, as presented by Math et al [8] often struggle with accuracy for large-scale or annual predictions due to climate change. Consequently, a more sophisticated weather prediction system is needed to help farmers determine optimal planting and harvesting timelines for rice, thereby preventing substantial losses. Singh et al. [9] highlighted the use of machine learning for irrigation scheduling and prediction, requiring weather data collected by appropriate sensors on the weather station.

**Table 1:** In depth analysis of Technology by using local weather forecasting for agriculture.

| | | Digital technologies | Weather parameters | Enhancement in Agriculture | References |
|---|---|---|---|---|---|
| **Data collection Technology** | low-cost sensor | air temperature and humidity sensor (DHT22), rain sensor (YL-83+LM393), atmospheric pressure sensor (BMP180) and LDR based intensity sensor | 1.Temeprature 2.Dew Point temperature 3.Relative humidity 4.Atmospheric Pressure 5.Solar irradiation 6.Diffuse Solar Irradiation 7.wind direction 8.wind speed 9.Rain accumulation 10.Rain duration 11.hail accumulation 12.hail duration 13.hail intensity 14.Precipitation 15.Soil temperature 16.Soil moisture | Weed Protection Warehousing Pest Protection Intelligent Spraying Decision Making Harvesting Soil Management Agriculture Supply Chain Irrigation | [22] |
| | | Node MCU, Radio circuit | | | [23] |
| | Short range sensor | ESP32, LoRaWAN, Sigfox, NB-IoT, Wi-Fi, Bluetooth, HW390 capacitive sensor | | | [8,9,24] |
| | long range sensor | Veml7700, NRF24L01, Long Range Radio (LoRa) protocol, Wireless Sensor- GPRS/3G/4G technology, | | | [9,22] |
| | data collection platform | low-cost WSN(wireless electronic modules), Thing speak, Zigbee | | | [23] |
| | | open-source Arduino, Thing Speak | | | [8,24] |
| | Secondary data | satellite remote sensing, Data from NASA, Data from IMD | | | [13] |
| | Autonomous data collection devices | Drones | | | [7] |
| | | GIS technology, remote sensing devices GPS based receivers | | | [13] |
| **Data storage layers** | block storage | MongoDB NoSQL database | | | [10] |
| | lower-level object storage | cloud computing, Amazons Elastic Computing Cloud (EC2) | | | [13,23] |
| | higher level object storage | Fog and Edge computing | | | [9] |
| | Object storage platform | Block chain, AWS | | | [26] |
| **Data Processing layers** | **Big Data analytics** — Artificial Intelligence | Support vector machine, Generative AI -Large Language Models (LLMs), ad boost, naïve bayes, k means clustering | | | [10,18] |
| | **Big Data analytics** — Machine learning | Random forest, linear regression, decision tree, | | | [2,18] |
| | Deep learning | Image recognition techniques, artificialneural network, gaussien mixture, | | | [11,25] |
| **Application layers** | UI component layer | a graphical user-friendly GUI, API | | | [5,8] |
| | UI processes layer | Watson Decision AI Platform | | | [18] |

Hachimi Chouaib et al. [10] proposed a system offering various services, including using machine learning to estimate reference evapo transpiration for effective irrigation, generating weather forecasts, and visualizing meteorological data. Kumar Rahul et al. [11] described the application of reinforcement learning techniques based on historical data to aid in weather forecasting. Overall, artificial intelligence in farming can be a powerful tool for weather forecasting, crop sustainability analysis, and pest and disease detection, as demonstrated by Singh Kanika et al. [9], who validated the accuracy of a trained model using a Random Forest regressor for producing weather summaries and crop advisories.

## 3. Smart -Local weather forecast and Agricultural enhancement

This paper examines the use of technology to improve local weather forecasting. Predicting local weather conditions can be challenging due to limited knowledge of localized phenomena. Often, we rely on secondary data from national weather forecasts. To address this, the paper discusses developing technologies to establish local meteorological stations for specific regions or communities. This requires exploring various technologies for localized weather forecasting. This paper proposes an innovative approach to develop a comprehensive ecosystem of IoT and AI-driven local weather forecasting for climate-smart agriculture. Leveraging the advancements in IoT technology, numerous cost-effective and energy-efficient sensor nodes can be deployed across farmlands to collect precise data on crops and local climate conditions. This data is then transmitted to a cloud based IoT framework via internet connectivity. The presented framework is designed to provide real-time data and AI-powered insights to researchers and farmers.

### 3.1 The theoretical Framework on IoT-AI research for Enhancing Local Weather forecast for smart agriculture.

It is structured around five key components: data acquisition, data storage, data processing, application, and decision-making. The proposed framework employs two primary data sources: locally gathered weather data and secondary data. The secondary data includes information from satellite, drone, and third-party providers like NASA, IMD etc. For the primary data, various IoT, environmental sensor, and weather monitoring techniques are used to collect local weather parameters such as rainfall, temperature, wind speed, wind direction, cloud cover, and light intensity. This data is acquired through diverse IoT devices and stored on platforms like Arduino, Thingspeak, and Zigbee. After collating the data from different locations, it is processed and archived in databases, including MongoDB, cloud computing platforms, edge computing, or Amazon Web Services (AWS). The master database compiled from local and secondary data sources is analyzed, categorized, and leveraged for predictive modeling. Diverse artificial intelligence and machine learning algorithms are explored to forecast local weather patterns and anticipated catastrophic events. A graphical user interface-based notification system is employed to effectively communicate more accurate predictions to local farmers, enabling improved management practices and cultivating a sustainable ecosystem. The proposed framework supports the intelligent stewardship of weather data to empower smart agricultural decision-making, optimizing operations from selection of crops to harvesting. The framework utilizes descriptive, diagnostic, and predictive data analysis. It employs AI-ML integration and data analytics to forecast future weather patterns, helping to mitigate severe events.

The framework consists of five key components as shown in the Fig. 1:

1. **Data Acquisition Layer**: This layer gathers diverse data from various IoT-based platforms and weather sensors. Secondary data is acquired from satellite and drone imagery, GIS information, and third-party meteorological services. Data communication platforms like WiFi, Bluetooth, DST sensors, and intelligent sensors are utilized.
2. **Data Storage Layer**: The collected data, primarily time series data, is stored using MongoDB and other platforms like cloud and edge computing. MongoDB is chosen for its scalability and suitability for handling large volumes of multisource data.
3. **Data Processing Layer**: This layer uses data analytics models, AI-ML fusion, and the data storage layer to extract insights and transform data into services. The data is classified and prepared for visualization.
4. **Application Layer**: Post-processing data is used to forecast various meteorological phenomena and impending disasters. A GUI notification system visualizes the data and alerts local farmers to early warnings. Rule-based approaches identify unique data patterns in time series, and notifications are sent via email and SMS.
5. **Decision Making Layer**: This final layer supports decision-making to prevent severe events based on the forecasting service's output. It provides parameters for predicting floods, droughts, human risks, environmental risks, significant temperature changes, and agricultural threats.

This integrated framework provides a comprehensive solution for local weather forecasting, benefiting both researchers and farmers while aligning with sustainable development objectives. From the standpoint of smart agriculture, the framework enables intelligent management of weather data to optimize practices like irrigation scheduling and crop selection. It employs historical data analysis to explain past events, diagnostic data analysis to identify anomalies, and predictive data analysis to forecast future weather conditions through the fusion of AI and machine learning techniques with data analytics.

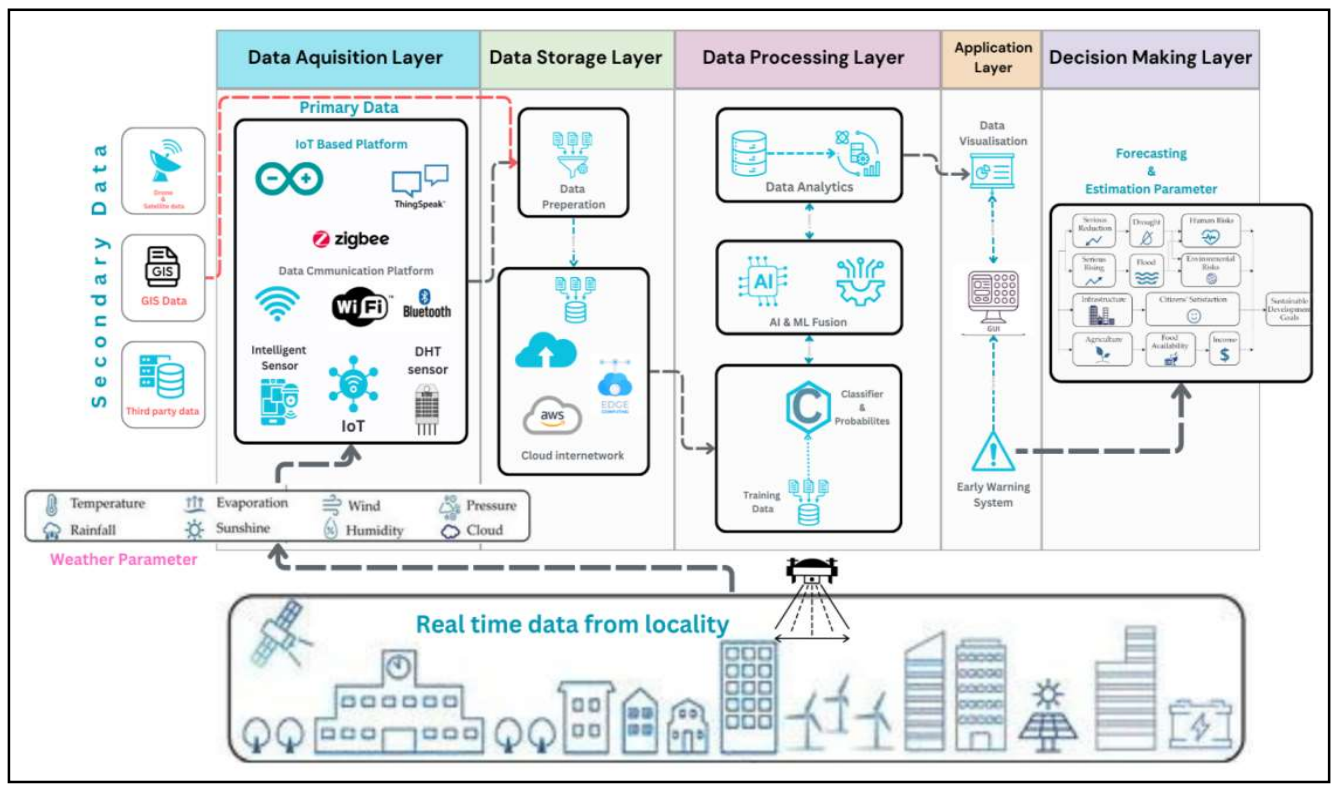

**Fig 1**: Local weather forecasting framework for smart agriculture using digital technologies.

## 4. Case study

The adoption of advanced agricultural practices has been driven by the emergence of digital technologies such as IoT, remote sensing capabilities, AI, and machine learning algorithms, in response to concerns about climate change. The existing literature has demonstrated the application of linear regression to model the relationships between weather variables using a linear equation. Furthermore, time-series forecasting techniques such as ARIMA and LSTM have been found to be well-suited for predicting variables like temperature and precipitation. Additionally, Support Vector Machines have proven to be a versatile approach, capable of handling both classification and regression tasks in the context of weather forecasting. Access to precise, real-time, and historical weather data for a specific location is crucial, as it enables farmers to optimize their sowing, irrigation, fertilization, and pest control practices. This can directly result in improved yields and crop quality. Farmers can gather and analyze local data on temperature, precipitation, soil structure, and other relevant crop-growing conditions using various IoT sensors as part of an agricultural weather monitoring system. Analyzing IoT data using machine learning algorithms allows for the identification of weather patterns and correlations, empowering proactive measures such as optimized irrigation and pest control. Consequently, agricultural enterprises can mitigate the disruptive effects of weather by taking proactive action, enhance farming and growth productivity, and reduce operational costs. Automated crop and weather monitoring reduces human error and enables more precise and efficient planting.

This section discusses and analyzes the various machine learning and statistical techniques leveraged by the organization to enhance weather forecasting for precision agriculture. The real-world case studies validate the effectiveness and applicability of the proposed framework. Each of these initiatives has unique strengths and capabilities that contribute to improving the accuracy and reliability of weather predictions, ultimately leading to enhanced agricultural productivity and resource optimization. This integrated case study explores in weather forecasting for precision agriculture across three distinct projects: Krishi Vidyapeeth Rahuri in Ahmednagar, India, the Meghdoot app in India, and IBM's AI innovations in agriculture. Despite differences in scope and scale, these initiatives share common objectives of enhancing agricultural productivity, improving disease detection, and optimizing resource utilization through precise weather forecasting and AI-driven insights [12].

### 4.1. Krishi Vidyapeeth Rahuri (KVR) Weather Station

The KVR weather station collected daily data on temperature, humidity, and rainfall over a ten-year period. This data is crucial for agricultural decision-making in the Ahmednagar region. The KVR station, located near major rivers Pravara and Mula, experiences significant annual rainfall, affecting local agriculture. They have used data acquisition and data processing layer as given in the above proposed framework (Fig. 1) to collect parameters such as humidity (%), temperature (°C), and rainfall (mm) to support local farming practices. This weather-based prediction and decision support system helps to manage the impact of seasonal changes on crop yield and resource allocation [13].

### 4.2. The Meghdoot app in India

The Meghdoot app which is a joint initiative of IMD (India Meteorological Department), IITM (Indian Institute of Tropical Meteorology) and ICAR (Indian Council of Agricultural Research). This app integrates observed weather data, forecasted data, and historical crop advisories to provide weather summaries and crop advisories for farmers. This case study focuses on two districts in Maharashtra (Gadchiroli) and one district in Punjab (Ludhiana). It involves compiling 307 past crop advisories for wheat and rice, along with a comprehensive set of observed weather data over 14 days. Weather parameters include temperature, relative humidity, wind speed, and rainfall as a part of data acquision layer had collected through IOT based platforms. Then data is stored in data storage layer and processed in data processing layer used as random forest regressor algorithm for predicting crop warnings and weather summaries, supported by OpenAI prompt engineering to enhance the model's natural language processing capabilities. Then data is visualized through an interface as given in application layer. The trained model is rigorously tested in the given theoretical proposed framework (Fig. 1) to ensure accuracy in weather summaries and crop advisories [2] by using the mobile app Meghdoot which is also a part of decision-making layer, through which farmers can access location specific weather based agro advisories over India to minimize crop losses in advance.

**4.3. IBM's AI Innovations in Agriculture**

IBM's AI technologies leverage IoT and precision agriculture to collect data from field sensors, livestock, and equipment. The sensors monitor soil moisture, temperature, humidity, and other environmental factors [14]. AI algorithms then use this data to create precise field maps, enabling customized planting, fertilization, and irrigation techniques [15]. This AI (Artificial Intelligence) & ML (Machine Learning) fusion (in Fig. 1) have followed the data processing layer where they have processed the collected weather parameters as secondary source then stored in storage layer and after processing, displayed the data at GUI (Graphical User Interface) in application layer and finally used for several applications for forecasting and early estimation of extreme weather events to obtain sustainable development goals (SDG). This ultimately leads to higher crop yields and more efficient resource utilization. IBM's AI also includes:

**Plant Disease Detection:** IBM's artificial intelligence algorithms can analyze crop photos to detect pest and disease indicators. These models employ machine learning and image recognition techniques to identify subtle changes in plant appearance that may signal the presence of illnesses. Farmers can take prompt action to stop the spread of diseases and minimize crop losses by identifying problems early [16].

**Crop Yield Prediction:** AI-driven crop yield prediction models utilize data from field sensors, drones, and satellites to estimate crop production potential. These models consider factors like soil quality, planting techniques, and weather patterns to provide insights into potential yields [17]. Farmers can use this information to plan better harvests, allocate resources, and enhance marketing strategies.

**Supply Chain Optimization:** IBM's AI solutions optimize the agricultural supply chain by analyzing data related to distribution, storage, and transportation. By predicting demand trends and optimizing routes, these services help to reduce waste, minimize transportation expenses, and ensure the timely delivery of fresh products to markets [18].

## 5. Conclusion

Digital advancements in artificial intelligence and machine learning present promising opportunities for the agricultural sector. These technologies enable continuous monitoring of environmental variables, soil conditions, and crop development, which can then be leveraged to enhance weather prediction accuracy and promote sustainable farming methods. The heightened efficiency and precision of weather monitoring will result in the rapid generation of vast data sets. Furthermore, Internet of Things weather stations that update weather information based on real-time sensor data have emerged as valuable tools across diverse domains, including climate research, agriculture, and disaster management. By using AI-powered methods, it may be possible to identify sudden climate changes earlier and take appropriate action. This paper explored the Integration of AI-ML and IoT techniques for weather prediction tasks to support agriculture, and the incorporation of IoT technologies that provide real-time, high-resolution meteorological data, representing a step forward. However, the lack of a comprehensive framework makes it difficult to fully implement this technology for creating a local weather forecasting system. This paper aims to understand the existing literature and provide a comprehensive overview of this technology using a case study approach. By developing a low-cost weather monitoring system using this framework, farmers could gain access to local, up-to-date climate information. This would allow farmers to complete their agricultural activities in a timely manner and mitigate the risk of economic losses. Furthermore, the platform could provide consumers and buyers with information about the supply chain, thereby enhancing the value proposition and potentially increasing farmer profitability.

## 6. Limitations and directions for future research:

Some key limitations of the study that should be addressed by researchers and practitioners include the lack of straightforward solutions, which poses a significant obstacle to the widespread adoption of AI in agriculture and local weather forecasting. More empirical research is required to improve these frameworks. Additionally, a lack of digital skills among farmers is a major challenge, as most farmers, particularly those in rural areas, lack the necessary technical expertise and awareness of these technologies [19]. Insufficient data collection methods are a crucial limitation for developing robust machine learning models, as the data infrastructure is a dynamic system [20].

Future research directions could explore how artificial intelligence has the potential to address diverse agricultural challenges, such as pest management and local weather forecasting. Developing new algorithms could support the monitoring of weeds, assessment of crop quality during harvesting and packaging, and identification of specific pest species through analysis of satellite imagery and historical data. Additionally, AI-powered systems could provide farmers with timely alerts on their mobile devices, enabling them to take appropriate action and utilize suitable pest control methods [21].

**Acknowledgements**: The authors sincerely acknowledge IMD Bhubaneswar for providing the required resources and permission for carrying out the present research for the conference presentation.

**Author Contributions**: Ms. PN: conceptualization, methodology, formal analysis, investigation, validation, writing original draft. Mr. SD: Framework conceptualization, resources, methodology supervision, conceptualization, validation, software, writing and editing.

**Declarations:** Competing Interests The authors declare that they have no known competing financial interests or personal relationships that could have appeared to influence the work reported in this paper.